# The vacuum state and minimum energy in Dirac's hole theory

Dan Solomon
Rauland-Borg Corporation
Mount Prospect, IL
Email: dan.solomon@rauland.com
Aug. 5, 2012

**Abstact.**

In Dirac's hole theory the vacuum state is assumed to be the state where all negative energy states are occupied and all positive energy states are unoccupied. This is often referred to as the Dirac sea. It is generally assumed that the Dirac sea is the minimum possible energy state. However it will be shown in this paper that this is not the case.

## 1. Introduction.

The solutions to the Dirac equation include states with both positive and negative energy. This creates a potential problem in that an electron in a positive energy state will tend to decay into a state with negative energy. This is not generally observed to occur. This objection can be overcome by assuming that all the negative energy states are occupied by a single electron. In this case the Pauli exclusion principle is used to prevent the decay of a positive energy electron into a negative energy state.

The *Dirac sea* is considered to be a system where all the negative energy states are occupied and all the positive energy states are unoccupied. It is generally assumed that the Dirac sea is the vacuum state of the system and the state of lowest energy. However, a number of papers by the author have shown that the latter assumption is not necessarily correct [1-4].

For example, Ref [1] examines a system in one space dimensional consisting of fermions with non-zero mass. The initial state of the system is that of the unperturbed Dirac sea. An electric potential is applied and then removed. The first order change in the energy of each negative energy electron in the Dirac sea can be calculated using perturbation theory. The total first order change in the energy of the Dirac sea can then be determined by summing over the change in the energy of each electron. It is shown that this change is negative. Therefore energy is extracted from the Dirac sea due to the perturbing potential.

Another example is given in Ref. [2] where massless fermions are considered. An electric field is applied and then removed. The change in energy of each electron can be calculated exactly for this situation. The total change in the energy of the system can be determined by summing up the change in the energy of each electron. It is shown that there is not a lower bound to the energy.

Ref [3] and [4] are based on work by Coutinho et al [5,6] where it was shown that Dirac's hole theory and quantum field theory can give different results. In [3] and [4] it was shown the reason for this is that Dirac's hole theory can have states with less energy than the vacuum state and quantum field theory can't.

In this paper we will add to the analysis of the previous papers. We will examine the Dirac sea for a system that consists of non-interacting massive electrons in 1-1 dimensional space-time. By non-interacting we mean that the electrons interact with an external potential but not with each other. We will postulate that the Dirac Sea is the lowest energy state and then show that this leads to a contradiction.

**2. Hole Theory**

In this section will discuss the basics of Dirac hole theory [7,8,9]. The effect of the assumption that all the negative energy states are occupied is to turn a one electron theory into an N-electron theory where $N \to \infty$. For a multiple electron theory the wave function is expressed as a Slater determinant,

$$\Psi^N(z_1, z_2, \ldots, z_N, t) = \frac{1}{\sqrt{N!}} \sum_P (-1)^s P(\psi_1(z_1, t)\psi_2(z_2, t)\cdots\psi_N(z_N, t)) \qquad (2.1)$$

where the $\psi_n(z,t)$ ($n = 1, 2, \ldots, N$) are orthogonal and normalized wave functions which obey the Dirac equation, P is a permutation operator which acts on the space coordinates, and s represent the number of interchanges in P.

Define the expectation value of a single particle operator $O_{op}(z)$ according to,

$$O_e = \int \psi^\dagger(z,t) O_{op}(z) \psi(z,t) dz \qquad (2.2)$$

In this equation $\psi(z,t)$ is a normalized single particle wave function. The N-electron operator is defined by,

$$O_{op}^N(z_1, z_2, ..., z_N) = \sum_{n=1}^{N} O_{op}(z_n) \tag{2.3}$$

This is the sum of one particle operators. The expectation value of the normalized N-electron wave function defined in (2.1) is,

$$O_e^N = \int \Psi^{N\dagger}(z_1, z_2, ..., x_N, t) O_{op}^N(z_1, z_2, ..., z_N) \Psi^N(z_1, z_2, ..., z_N, t) dz_1 dz_2 ... dz_N \tag{2.4}$$

This can be demonstrated to be equal to,

$$O_e^N = \sum_{n=1}^{N} \int \psi_n^\dagger(z,t) O_{op}(z) \psi_n(z,t) dz \tag{2.5}$$

The N-electron expectation value is simply the sum of the single particle expectation values that are associated with each individual wave functions $\psi_n$. We will write this symbolically as,

$$\langle \Psi^N O_e \Psi^N \rangle = \sum_{n=1}^{N} \int \psi_n^\dagger(z,t) O_{op}(z) \psi_n(z,t) dz \tag{2.6}$$

## 3. Assumption of minimum energy.

Each of the wave functions $\psi_n(z,t)$ in the Slater determinant of Eq. (2.1) obeys the Dirac equation. In the present discussion we will consider the Dirac equation in one space dimension in the presence of a static inverse square well potential $V_\eta(z)$ where $V_\eta(z)$ is a given by,

$$V_\eta(z) = \begin{cases} 0 \text{ if } |z| > a/2 \\ -\eta \text{ if } |z| < a/2 \end{cases} \tag{3.1}$$

where $\eta \geq 0$. The Hamiltonian operator is,

$$H_\eta = H_0 + V_\eta(z) \tag{3.2}$$

where,

$$H_0 = -i\sigma_1 \frac{\partial}{\partial z} + \sigma_3 m \tag{3.3}$$

and where $m$ is the electron mass and $\sigma_j$ are the Pauli Matrices.

The eigenfunction solutions $\psi_k(\eta;z)$ satisfy,

$$E\psi_k(\eta;z) = (H_0 + V_\eta(z))\psi_k(\eta;z) \tag{3.4}$$

A detailed analysis of this problem is given by Dehghan and Gousheh [10]. According to [10] there are a continuum of negative solutions for $E = -\sqrt{k^2 + m}$ and a continuum of positive energy solutions for $E = +\sqrt{k^2 + m}$ where $k$ varies continuously from 0 to $\infty$. If $\eta \ne 0$ there are also a finite number of bound state solutions. If $m \ge \eta > 0$ the energy of the bound state solutions are all positive and satisfy $m > E_b > 0$. If $2m > \eta > m$ there may exist negative energy bound state solutions with $0 > E_b > -m$. In the following we will assume that $m \ge \eta \ge 0$ so that there are no negative energy bound states. Following [10] we will designate the solutions in the negative energy continuum by $v_{k,j}(\eta, z)$ and solutions in the positive energy continuum solutions by $u_{k,j}(\eta, z)$ where $j = \pm$ is the parity. The bound state solution are designated by $\chi_{b,j}(\eta, z)$. Free field solutions are obtained by taking $\eta = 0$ for $v_{k,j}(\eta, z)$ and $u_{k,j}(\eta, z)$. For the free field there are no bound solutions.

Recall that the *Dirac sea* is the state in which all negative energy states are occupied by a single electron and all positive energy states are unoccupied. For a given $\eta$ (with $m > \eta > 0$) the wave function for the Dirac sea is the Slater determinant $\Psi_\eta^{ds}$ which consists of all negative energy wave functions $v_{k,j}(\eta, z)$. The energy of the Dirac sea is then given as,

$$\left\langle \Psi_\eta^{ds} H_\eta \Psi_\eta^{ds} \right\rangle = \sum_{j=\pm} \int_0^\infty dk \int v_{k,j}^\dagger(\eta, z) H_\eta v_{k,j}(\eta, z) dz \qquad (3.5)$$

The question we want to address is whether or not the Dirac sea is the lowest energy state. If this is the case then the energy of any other state must be greater. If this were true this would mean that,

$$\left\langle \Psi_\alpha^{ds} H_\eta \Psi_\alpha^{ds} \right\rangle > \left\langle \Psi_\eta^{ds} H_\eta \Psi_\eta^{ds} \right\rangle \text{ if } \eta \ne \alpha \qquad (3.6)$$

where,

$$\left\langle \Psi_\alpha^{ds} H_\eta \Psi_\alpha^{ds} \right\rangle = \sum_{j=\pm} \int_0^\infty dk \int v_{k,j}^\dagger(\alpha, z) H_\eta v_{k,j}(\alpha, z) dz \qquad (3.7)$$

It also follows that,

$$\langle \Psi_\eta^{ds} H_\alpha \Psi_\eta^{ds} \rangle > \langle \Psi_\alpha^{ds} H_\alpha \Psi_\alpha^{ds} \rangle \tag{3.8}$$

Let $\alpha = 0$. Use this along with (3.2) in the above to obtain,

$$\langle \Psi_0^{ds} (H_0 + V_\eta) \Psi_0^{ds} \rangle > \langle \Psi_\eta^{ds} (H_0 + V_\eta) \Psi_\eta^{ds} \rangle \tag{3.9}$$

and,

$$\langle \Psi_\eta^{ds} H_0 \Psi_\eta^{ds} \rangle > \langle \Psi_0^{ds} H_0 \Psi_0^{ds} \rangle \tag{3.10}$$

Rearrange terms in (3.9) to obtain,

$$\langle \Psi_0^{ds} H_0 \Psi_0^{ds} \rangle - \langle \Psi_\eta^{ds} H_0 \Psi_\eta^{ds} \rangle > \langle \Psi_\eta^{ds} V_\eta \Psi_\eta^{ds} \rangle - \langle \Psi_0^{ds} V_\eta \Psi_0^{ds} \rangle \tag{3.11}$$

Referring to (3.10) the left side of the above must be negative. Therefore a necessary condition for both (3.10) and (3.11) to be true is,

$$0 > \langle \Psi_\eta^{ds} V_\eta \Psi_\eta^{ds} \rangle - \langle \Psi_0^{ds} V_\eta \Psi_0^{ds} \rangle \tag{3.12}$$

Using (3.1) this yields

$$0 > -\eta \left( \sum_{j=\pm} \int_0^\infty \frac{dk}{2\pi} \int_{-a/2}^{+a/2} v_{k,j}^\dagger(\eta, z) v_{k,j}(\eta, z) dz - \sum_{j=\pm} \int_0^\infty \frac{dk}{2\pi} \int_{-a/2}^{+a/2} v_{k,j}^\dagger(0, z) v_{k,j}(0, z) dz \right) \tag{3.13}$$

There is a potential problem with evaluating this expression in that the two terms, $\langle \Psi_\eta V_\eta \Psi_\eta \rangle$ and $\langle \Psi_0 V_\eta \Psi_0 \rangle$ are both infinite so we are subtracting one infinity from another. To evaluate this we will follow the approach of Ref [10] and rearrange terms to obtain,

$$0 > -\eta \sum_{j=\pm} \int_0^\infty \frac{dk}{2\pi} \int_{-a/2}^{+a/2} \left( v_{k,j}^\dagger(\eta, z) v_{k,j}(\eta, z) - v_{k,j}^\dagger(0, z) v_{k,j}(0, z) \right) dz \tag{3.14}$$

The integrand will fall off sufficiently fast as $k \to \infty$ so that the above quantity is finite and readily evaluated by numerical methods.

The net charge density $\Delta\rho(\eta, z)$ is defined as,

$$\Delta\rho(\eta, z) = \sum_{j=\pm} \int_0^\infty \frac{dk}{2\pi} \left( v_{k,j}^\dagger(\eta, z) v_{k,j}(\eta, z) - v_{k,j}^\dagger(0, z) v_{k,j}(0, z) \right) \tag{3.15}$$

This is the vacuum polarization charge. It represents the difference between the vacuum charge when the potential is applied and the vacuum charge when the potential is absent. Note that in the context of this paper the charge density represents the density of electrons. To get the *electrical* charge density one would have to multiple $\Delta\rho(\eta,z)$ by the charge on the electron.

Use (3.15) in (3.14) to obtain,

$$0 > -\eta Q(\eta) \tag{3.16}$$

where,

$$Q(\eta) = \int_{-a/2}^{+a/2} \Delta\rho(\eta,z)\,dz \tag{3.17}$$

$Q(\eta)$ is the net charge within the square well, that is, in the region $a/2 > z > -a/2$. As will be shown in the next section $Q(\eta)$ is negative. Therefore, since $\eta$ is positive, (3.16) is false. Therefore the relationships (3.8) and (3.10) can't both be true and the assumption that the Dirac sea is the lowest energy state leads to a contradiction and is false.

## 4. Charge density

In this section we will determine the net charge $Q(\eta)$. It is the change in the number of electrons in the region $|z| < a/2$ due to the application of the potential $V_\eta(z)$ defined in (3.1).

The solutions to the Dirac equation for the inverse square potential are given in [10]. In order to determine $Q(\eta)$ we need the solutions for the negative energy states in the region $a/2 > z > -a/2$. Ref. [10] divides the negative energy continuum solutions $v_{k,\pm}(\eta,z)$ into two types. These are $v_{1k,\pm}(\eta,z)$ and $v_{2k,\pm}(\eta,z)$. $v_{2k,\pm}(\eta,z)$ is the solution for $-m-\eta \leq -\sqrt{k^2+m^2} \leq -m$ and $v_{1k,\pm}(\eta,z)$ are the solutions for $-\sqrt{k^2+m^2} \leq -m-\eta$. From [10] these are,

$$v_{1k,\pm}(\eta,z) = d_\pm \begin{pmatrix} e^{ipz} \pm e^{-ipz} \\ \dfrac{p}{E+\eta+m}\left(e^{ipz} \mp e^{-ipz}\right) \end{pmatrix} \quad \text{for } -\frac{a}{2} \leq z \leq \frac{a}{2} \tag{4.1}$$

and

$$v_{2k,\pm}(\eta,z) = g_{\pm}\begin{pmatrix} e^{-\lambda z} \pm e^{+\lambda z} \\ \dfrac{i\lambda}{E+\eta+m}\left(e^{-\lambda z} \mp e^{+\lambda z}\right) \end{pmatrix} \text{ for } -\dfrac{a}{2} \leq z \leq \dfrac{a}{2} \qquad (4.2)$$

where $E = -\sqrt{k^2 + m^2}$, $p = \sqrt{(E+\eta)^2 - m^2}$, $\lambda = \sqrt{m^2 - (E+\eta)^2}$

$$d_{\pm} = \left[\dfrac{E+m}{E\left(2 \pm e^{ipa} \pm e^{-ipa} + \dfrac{2 \mp e^{ipa} \mp e^{-ipa}}{\gamma^2}\right)}\right]^{\frac{1}{2}} \quad ; \quad \gamma = \dfrac{k(E+\eta+m)}{p(E+m)} \qquad (4.3)$$

$$g_+ = N_{2+}; \quad g_- = iN_{2-} \qquad (4.4)$$

$$N_{2\pm} = \left[\dfrac{E+m}{E\left(\pm 2 + e^{\lambda a} + e^{-\lambda a} + \dfrac{\mp 2 + e^{\lambda a} + e^{-\lambda a}}{\gamma'^2}\right)}\right]^{\frac{1}{2}} \quad ; \quad \gamma' = \dfrac{k(E+\eta+m)}{\lambda(E+m)} \qquad (4.5)$$

Define the charge density for a given mode as

$$\rho_{1k,\pm}(\eta,z) = v^{\dagger}_{1k,\pm}(\eta,z)v_{1k,\pm}(\eta,z) \text{ for } -\sqrt{k^2+m^2} < -m-\eta \qquad (4.6)$$

and,

$$\rho_{2k,\pm}(\eta,z) = v^{\dagger}_{2k,\pm}(\eta,z)v_{2k,\pm}(\eta,z) \text{ for } -m-\eta \leq -\sqrt{k^2+m^2} \leq -m \qquad (4.7)$$

Using the previous relationships these can be evaluated as,

$$\rho_{1k,\pm}(\eta,z) = 2|d_{\pm}|^2\left[(1 \pm \cos(2pz)) + \left(\dfrac{p^2}{(E+\eta+m)^2}\right)(1 \mp \cos(2pz))\right] \qquad (4.8)$$

$$\rho_{2k,\pm}(\eta,z) = 2|g_{\pm}|^2\left[(\pm 1 + \cosh(2\kappa z)) + \left(\dfrac{\lambda^2}{(E+\eta+m)^2}\right)(\cosh(2pz) \mp 1)\right] \qquad (4.9)$$

For $\eta = 0$ we define,

$$\rho_{k,\pm}(0,z) = v^{\dagger}_{1k,\pm}(0,z)v_{1k,\pm}(0,z) \qquad (4.10)$$

From the above define,

$$\Delta\rho_{2k,\pm}(\eta,z) = \rho_{2k,\pm}(\eta,z) - \rho_{k,\pm}(0,z) \tag{4.11}$$

and,

$$\Delta\rho_{1k,\pm}(\eta,z) = \rho_{1k,\pm}(\eta,z) - \rho_{k,\pm}(0,z) \tag{4.12}$$

These are the change in the charge density for a given mode due to due the application of the potential. Also define,

$$\Delta\rho_{\pm}(\eta,z) = \int_0^{k_1} \frac{dk}{2\pi} \Delta\rho_{2k,\pm}(\eta,z) + \int_{k_1}^{\infty} \frac{dk}{2\pi} \Delta\rho_{1k,\pm}(\eta,z) \tag{4.13}$$

where $k_1 = \sqrt{2m\eta + \eta^2}$. Therefore the net charge density in the region $a/2 > |z|$ is,

$$\Delta\rho(\eta,z) = \Delta\rho_{+}(\eta,z) + \Delta\rho_{-}(\eta,z) \tag{4.14}$$

Define,

$$Q_{2k,\pm}(\eta) = \int_{-a/2}^{+a/2} \Delta\rho_{2k,\pm}(\eta,z)dz \; ; \; Q_{1k,\pm}(\eta) = \int_{-a/2}^{+a/2} \Delta\rho_{1k,\pm}(\eta,z)dz \tag{4.15}$$

and,

$$Q_{\pm}(\eta) = \int_0^{k_1} \frac{dk}{2\pi} Q_{2k,\pm}(\eta) + \int_{k_1}^{\infty} \frac{dk}{2\pi} Q_{1k,\pm}(\eta) \tag{4.16}$$

Therefore the total net charge in the inverse square well (the region $|z| < a/2$) is,

$$Q(\eta) = Q_{+}(\eta) + Q_{-}(\eta) \tag{4.17}$$

The above expressions can be evaluated numerically. The following table on the next page gives $Q(\eta)$ for different values of the potential $\eta$ and width $a$. In all case $m=1$. The result is that $Q(\eta)$ in negative for all values in this table. This means that the region $|z| < a/2$ becomes depleted of electrons as $\eta$ increase from zero. Therefore the change in the electric charge within the region $|z| < a/2$ would be positive because electrons carry a negative charge.

| a | $\eta$ | $Q(\eta)$ |
|---|---|---|
| 1 | 1/10 | -0.021 |
| 1 | 1/2 | -0.103 |
| 1 | 1 | -0.204 |
| 5 | 1/10 | -0.147 |
| 5 | 1/2 | -0.733 |
| 5 | 1 | -1.46 |
| 10 | 1/10 | -0.306 |
| 10 | 1/2 | -1.53 |
| 10 | 1 | -3.05 |

In conclusion we have found that the assumption that the Dirac Sea is a state of minimum energy leads to a contradiction and cannot be true. This is consistent with previous results [1-4] which show that the Dirac Sea is not a lower bound to the energy in Dirac hole theory.

**References.**